\documentclass[epj]{webofc}
\usepackage[varg]{txfonts}   % Web of Conferences font
\newcommand{\be}{\begin{equation}}
\newcommand{\ee}{\end{equation}}
\newcommand{\ma}{\mathcal{A}}
\newcommand{\Tr}{{\rm Tr}}
\woctitle{CNR*13}
\begin{document}
\title{Recent Advances in the Microscopic Calculations of Level Densities by the Shell Model Monte Carlo Method}

\author{Y. Alhassid\inst{1}\fnsep\thanks{\email{yoram.alhassid@yale.edu}} \and M. Bonett-Matiz\inst{1}
\and S. Liu \and A. Mukherjee\inst{2} \and H. Nakada\inst{3}}

\institute{Center for Theoretical Physics, Sloane Physics Laboratory,Yale University, New Haven, CT 06520, USA
\and  ECT*, Villa Tambosi, I-38123 Villazzano, Trento, Italy
\and  Department of Physics, Graduate School of Science,  Chiba University, Inage,  Chiba 263-8522, Japan
}

\abstract{%
The shell model Monte Carlo (SMMC) method enables calculations in model spaces that are many orders of magnitude larger than those that can be treated by conventional methods, and is particularly suitable for the calculation of level densities in the presence of correlations. We review recent advances and applications of SMMC for the microscopic calculation of level densities. Recent developments include (i) a method to calculate accurately the ground-state energy of an odd-mass nucleus, circumventing a sign problem that originates in the projection on an odd number of particles, and (ii) a method to calculate directly level densities, which, unlike state densities, do not include the spin degeneracy of the levels. We calculated the level densities of a family of nickel isotopes $^{59-64}$Ni and of a heavy deformed rare-earth nucleus $^{162}$Dy and found them to be in close agreement with various experimental data sets.
}
\maketitle
\section{Introduction}
\label{intro}

The level density is required in the calculation of transition rates through Fermi's golden rule and has an important role in the Hauser-Feshbach theory of statistical nuclear reactions~\cite{Hauser1952}. Its calculation in the presence of interactions is a challenging many-body problem. A suitable framework that takes into account correlations and shell effects is the configuration-interaction (CI) shell model. However, the combinatorial growth of the dimension of the many-particle model space with the number of valence orbitals and/or the number of nucleons hinders application of the CI shell model in mid-mass and heavy nuclei. This problem can be overcome in part through the use of the shell model Monte Carlo (SMMC) method~\cite{Lang1993,Alhassid1994,Koonin1997,Alhassid2001}. In contrast to conventional diagonalization methods, the SMMC method scales much more gently with the dimension of the single-particle model space and enables calculations in much larger many-particle model spaces. As a finite-temperature method, it is particularly suitable for the calculation of statistical nuclear properties such as the level density. SMMC nuclear state densities were calculated in mid-mass~\cite{Nakada1997,Langanke1998,Alhassid1999}  and heavy~\cite{Alhassid2008,Ozen2013} nuclei, and were found to be in good agreement with various experimental data sets. Projection methods were successfully implemented in SMMC to determine the parity~\cite{Nakada1997,Alhassid2000,Ozen2007} and spin~\cite{Alhassid2007} distributions of level densities.

Quantum Monte Carlo methods for fermions often suffer from a sign problem that leads to large statistical errors. The dominant collective components~\cite{Zuker1996} of realistic effective nuclear interactions have a good Monte Carlo sign. The smaller bad-sign components can be treated using the extrapolation method of Ref.~\cite{Alhassid1994}.  Good-sign interactions are often sufficient for realistic calculations of nuclear densities. These good-sign interactions are free of the sign problem in the grand-canonical ensemble, in which the number of particles fluctuates. However, in the finite nucleus, it is important to use the canonical ensemble, in which both the proton and neutron numbers are fixed. While the projection on an even number of particles keeps the sign good, the projection on an odd number of particles leads to a new sign problem that is mild at intermediate temperatures but becomes severe at low temperatures.  Consequently,  applications of SMMC to odd-mass (and odd-odd) nuclei have been hampered by this odd-particle sign problem.

Recently, we introduced a method to calculate accurately the ground-state energy of an odd-particle system, circumventing the odd-particle sign problem, and we applied it to nuclei in the iron region~\cite{Mukherjee2012}.  This made possible the first accurate SMMC calculations of densities of odd-mass nuclei~\cite{Bonett2013}.

In canonical SMMC calculations, a trace is taken over the complete Hilbert space for a fixed number of protons and neutrons. Thus the calculated density is the {\em state} density, which includes the $2J+1$ magnetic degeneracy of each level with spin $J$. However, the experimentally measured density is often the {\em level} density, in which each level is counted just once, irrespective of its magnetic degeneracy.

In Ref.~\cite{Alhassid2007} we introduced a spin projection method in SMMC to determine the spin distribution $\rho_J(E_x)$ of nuclear levels at excitation energy $E_x$. The level density is then given by $\tilde\rho(E_x) = \sum_J \rho_J(E_x)$. However, the statistical errors of $\rho_J(E_x)$ increase with $J$ and make such a calculation of the level density impractical. Recently, we showed that a projection on the minimal absolute value of the magnetic quantum number can be used to calculate directly accurate level densities in SMMC~\cite{Alhassid2013}. Here we review the application of this method for calculating the level densities of a family of nickel isotopes $^{59-64}$Ni~\cite{Bonett2013}.  We find close agreement with level densities extracted from recent measurements of proton evaporation spectra~\cite{Voinov2012} and with level counting data. We also calculated the level density of a typical heavy deformed rare-earth nucleus $^{162}$Dy and compared it with available experimental data~\cite{Alhassid2013}.

\section{State densities by the shell model Monte Carlo (SMMC) method}

In this section, we briefly discuss the SMMC method and its application to the calculations of state densities.

\subsection{SMMC}
\label{SMMC}

The Gibbs operator $e^{-\beta H}$, describing a many-particle system with a Hamiltonian $H$ at inverse temperature $\beta=1/T$,  can be thought of as a propagator in imaginary time $\beta$. The SMMC method is based on the Hubbard-Stratonovich (HS) transformation~\cite{HS-trans}, in which the propagator $e^{-\beta H}$ is written as a superposition of propagators of  non-interacting particles moving in external auxiliary fields that depend on the imaginary time $\tau$ ($0 \leq \tau \leq \beta$). Formally, the HS transformation can be written as a functional integral
\be\label{HS}
e^{-\beta H} = \int D[\sigma] G_\sigma U_\sigma
\ee
over the auxiliary fields $\sigma$, where $G_\sigma$ is a Gaussian weight and $U_\sigma$ is a one-body propagator of non-interacting particles moving in external fields $\sigma=\sigma(\tau)$.

The expectation value of an observable $O$ at finite temperature $T=1/\beta$ can be written as
\be \label{observable}
\langle O\rangle = {\Tr \,( O e^{-\beta H})\over  \Tr\, (e^{-\beta H})} = {\int D[\sigma] W_\sigma \Phi_\sigma \langle O \rangle_\sigma
\over \int D[\sigma] W_\sigma \Phi_\sigma} \;,
\ee
where $W_\sigma = G_\sigma |\Tr\, U_\sigma|$
 is a positive-definite function, $\Phi_\sigma = \Tr\, U_\sigma/|\Tr\, U_\sigma|$ is the Monte Carlo sign, and $\langle O \rangle_\sigma =
 {\rm Tr} \,( O U_\sigma)/ {\rm Tr}\,U_\sigma$  is the thermal expectation value of the observable for a given configuration of the auxiliary fields $\sigma$.

  Since $U_\sigma$ is a one-body propagator it can be represented in the single-particle space by an $N_s\times N_s$ matrix ${\bf U}_\sigma$, where $N_s$ is the number of single-particle states.
 Quantities that appear in the integrands of Eq.~(\ref{observable}) can be expressed in terms of ${\bf U}_\sigma$. For example, the grand-canonical many-particle trace of $U_\sigma$ is given by
 \be
 {\rm Tr}\; U_\sigma = \det ( {\bf 1} + {\bf U}_\sigma)\;.
 \ee

 In the finite nucleus it is important to use the canonical ensemble, in which the number of protons and the number of neutrons are fixed. Thus, in Eq.~(\ref{observable}) we take the traces at fixed number of protons and neutrons using a discrete Fourier sum for each of the particle-number projections.

  In practical calculations, the time is discretized using a finite time slice $\Delta\beta$ and the results are extrapolated to $\Delta\beta=0$. The integration over the large number of auxiliary fields is done by Monte Carlo methods.  We approximate the integral over each of the auxiliary fields at any given time slice by a three-point quadrature formula, and use importance sampling to select  uncorrelated configurations $\sigma_k$ of the auxiliary fields that are distributed according to the positive-definite distribution $W_\sigma$. The expectation value in Eq.~(\ref{observable}) is then estimated from
  \be\label{MC-average}
\langle O\rangle \approx  {\sum_k
  \langle  O \rangle_{\sigma_k} \Phi_{\sigma_k} \over \sum_k \Phi_{\sigma_k}} \;.
\ee

\subsection{State densities}
\label{state_densities}
The state density $\rho(E)$ is related to the canonical partition function $Z(\beta)=\Tr e^{-\beta H}$ by an inverse Laplace transform
\be\label{Laplace}
\rho(E) ={1 \over 2\pi i} \int_{-i\infty}^{i\infty} d\beta\, e^{\beta E} Z(\beta) \;.
\ee
We evaluate the integral in (\ref{Laplace}) using the saddle-point approximation to obtain an average state density
\be
\rho(E) \approx {1 \over \sqrt{2\pi T^2 C}} e^{S(E)} \;,
\ee
 where $S$ is the canonical entropy and $C$ is the canonical heat capacity. In SMMC, we calculate the thermal energy as the expectation value of the Hamiltonian $E(\beta)=\langle H\rangle$, and then integrate the thermodynamic relation $-\partial \ln Z/\partial \beta=E(\beta)$ to find the partition function $Z(\beta)$. The entropy and heat capacity are then calculated from
\be
 S =\ln Z + \beta E \;;\;\;\; C=\frac{dE}{dT}=-\beta^2 {dE \over d\beta} \;.
\ee
 The derivative in the expression for the heat capacity is carried out numerically. This typically leads to large statistical errors at low excitation energies. These errors are reduced significantly by using the same set of auxiliary fields to calculate both $E(\beta + \delta\beta)$ and  $E(\beta - \delta\beta)$ in the numerical derivative, and taking into account correlated errors~\cite{Liu2001}.

\section{Odd-particle systems}
\label{Green_functions}
The odd-particle sign problem has hindered accurate calculations of the ground-state energy of the odd-particle system. In this section we review a Green's function method that we recently introduced to circumvent this sign problem~\cite{Mukherjee2012}.

 Consider an even-even nucleus $\ma=(Z,N)$ with $Z$ protons and $N$ neutrons. We define its imaginary-time scalar single-particle Green's functions by
\be\label{green}
G_{\nu}(\tau) = \frac{\Tr_{\ma}\left[~e^{-\beta H} \mathcal{T} \sum_m a_{\nu m}(\tau) a^\dagger_{\nu m}(0)\right]}{\Tr_{\ma}~e^{-\beta H}}\;.
\ee
 Here $\nu \equiv (n l j)$ denotes the single-particle orbital of a nucleon with radial quantum number $n$, orbital angular momentum $l$ and total spin $j$, $\mathcal{T}$ denotes
time ordering and $a_{\nu m}(\tau)\equiv e^{\tau H} a_{\nu m} e^{-\tau H}$ (for $-\beta \leq \tau\leq \beta$) is an annihilation operator of a nucleon at imaginary time $\tau$ in a
single-particle with orbital $\nu$ and magnetic quantum number $m$.

The Green's functions in (\ref{green}) connect between states of the even-even nucleus and the corresponding  odd-mass nucleus. For large values of $\beta$, the Green's function $G_\nu$ with $\nu=n l j$ connects the $J=0$ ground state of the even-even nucleus to states with $J=j$ in the corresponding odd-mass nucleus.  The asymptotic~\cite{Mukherjee2012} form of the Green's functions is given by
\be
G_{\nu}(\tau) \sim e^{- \Delta E_{J=j}(\ma_{\pm}) |\tau|} \;,
\ee
where $\ma_\pm$ denotes the even-odd nuclei $(Z,N \pm 1)$ when $\nu$ is a neutron orbital, and the odd-even nuclei $(Z \pm 1,N)$ when $\nu$ is a proton orbital. The $+$ ($-$) subscript should be used for $\tau > 0$  ($\tau \leq 0$). Here $\Delta E_{J=j}(\ma_{\pm})$ is the energy difference between the lowest-energy state with spin $J$ of the $\ma_{\pm}$-particle nucleus and the ground state of the $\ma$-particle nucleus. We find $\Delta E_{J=j}(\ma_{\pm})$ from the slope of a straight line fit to $|\ln G_\nu|$. We can then determine $E_{J=j}(\ma_{\pm})$ of the corresponding odd-mass nucleus $\ma\pm$  since the ground-state energy of the even-even nucleus $\ma$ can be calculated accurately in  direct SMMC calculations. The ground-state energy of the odd-mass nucleus is determined as the lowest energy $E_{J=j}(\ma_{\pm})$ among all possible $J$ values.

\begin{figure}[h!]
\centering
\sidecaption
\includegraphics[width=0.5\columnwidth,clip]{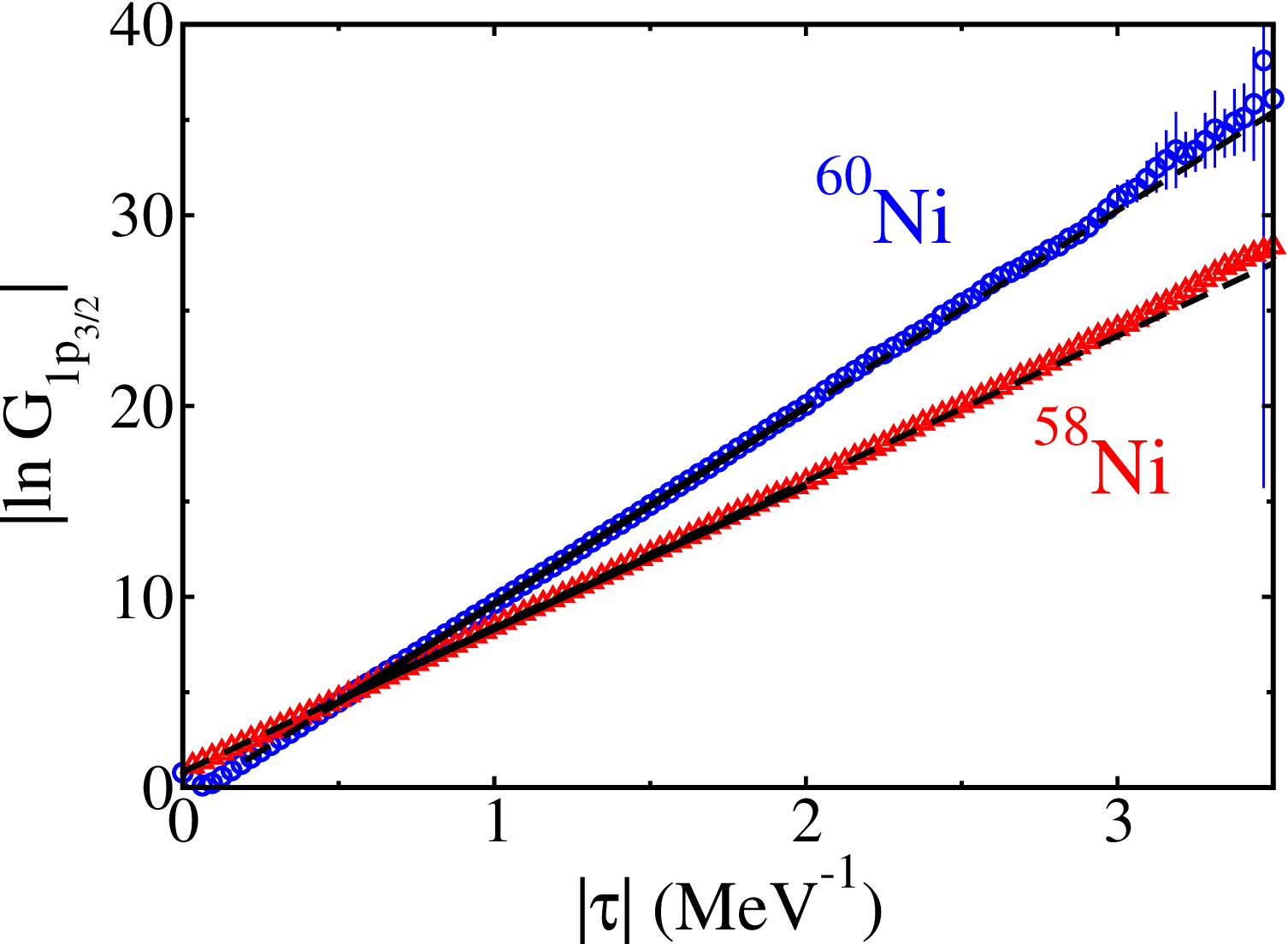}
\caption{$|\ln G_{1 p_{3/2}}|$ versus $|\tau|$ for $^{58}$Ni (triangles) and $^{60}$Ni (circles) at $\beta=4$  MeV$^{-1}$ (using a time slice of $\Delta \beta=1/32$  MeV$^{-1}$). The lines are straight line fits in the asymptotic interval $0.5 \leq \tau \leq 2$ (in units of MeV$^{-1}$), and their slopes determine the energy differences $\Delta E_{J=3/2}(^{59}{\rm Ni})$ between the lowest $J=3/2$ state in $^{59}$Ni and the ground state of $^{58}$Ni or $^{60}$Ni.}
\label{Ni_green}       
\end{figure}

We demonstrate our method for the odd-mass nucleus $^{59}$Ni, whose ground-state energy can be extracted using independently the Green's functions of $^{58}$Ni and $^{60}$Ni.  In Fig.~\ref{Ni_green} we show $|\ln G_\nu|$ as a function of $|\tau|$ for the $1p_{3/2}$ neutron orbital for $^{58}$Ni (triangles) and $^{60}$Ni (circles) at $\beta=4$ MeV$^{-1}$. The lines are straight line fits whose slopes determine the respective energy differences $\Delta E_{J=3/2}(^{59}{\rm Ni})$.

\begin{figure}[h!]
\centering
\sidecaption
\includegraphics[width=0.5\columnwidth,clip]{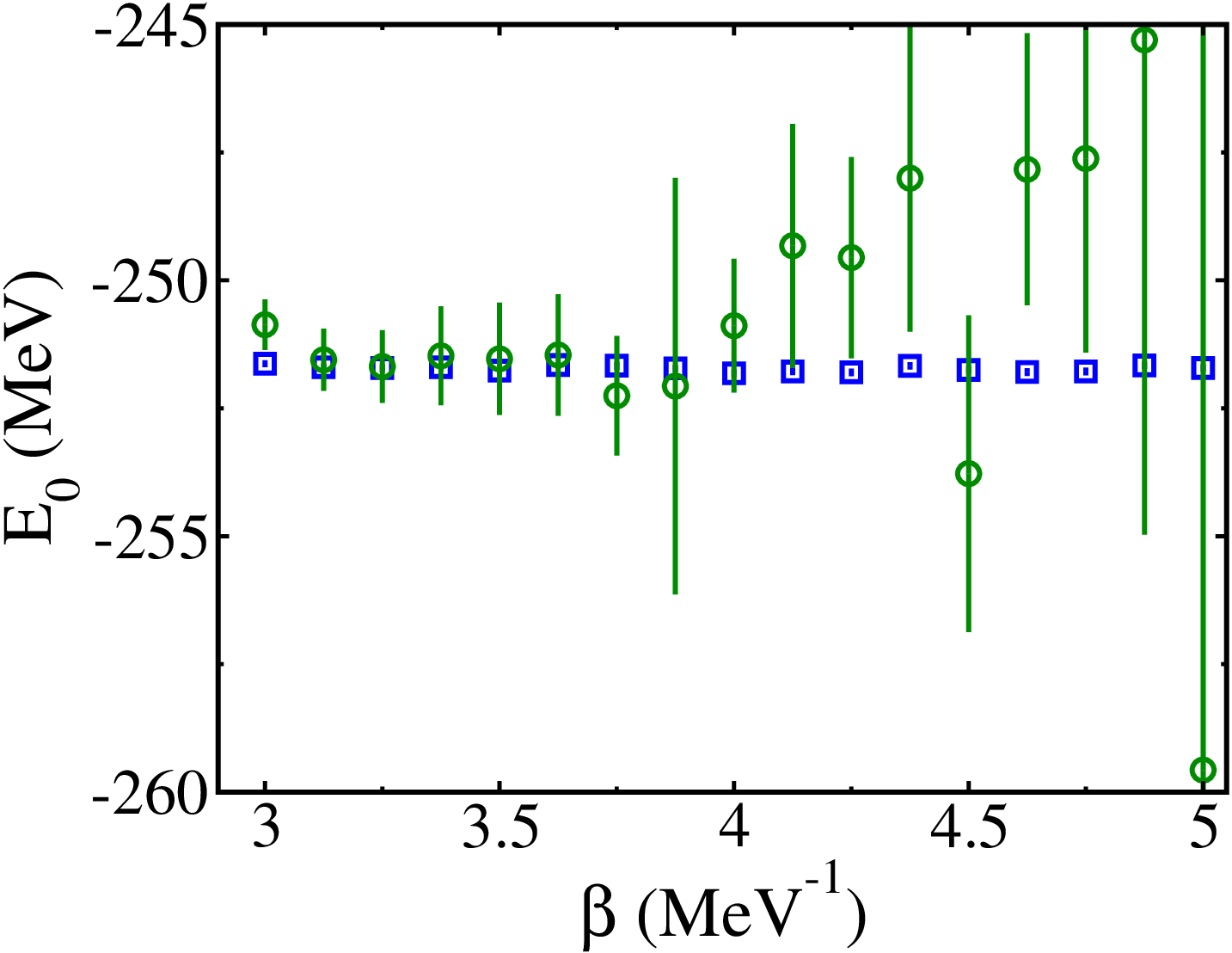}
\caption{The ground-state energy $E_0$ of the odd-mass nucleus $^{59}$Ni versus $\beta$. The Green's function method results (squares) are compared with direct SMMC calculations (circles). The odd-particle sign problem leads to large statistical errors in the direct calculations that increase with $\beta$. The error bars of the energies obtained in the Green's function method are smaller than the size of the symbols.}
\label{Ni59_gs_comp}      
\end{figure}

In Fig.~\ref{Ni59_gs_comp} we compare the ground-state energy of $^{59}$Ni that is extracted from the Green's function method (squares) with direct SMMC calculations in $^{59}$Ni (circles)
  using a time slice of $\Delta\beta=1/32$ MeV$^{-1}$. In the Green's function method we take an average of the energies obtained from the two even-mass nuclei $^{58}$Ni and $^{60}$Ni. We observe that the statistical errors in the Green's function method are much smaller than the those in the direct calculations, and are comparable to typical statistical errors obtained for even-even nuclei.

\section{Level densities in SMMC}
\label{level_densities}

In Sect.~\ref{state_densities} we discussed the SMMC calculation of the state density, which includes the $2J+1$  magnetic degeneracy of each level with spin $J$. However, experiments often measure the level density, which do not include the spin degeneracy of the levels. In this section we discuss a recent method to calculate directly the level density in SMMC~\cite{Alhassid2013}.

We observe that each level with spin $J$ has exactly one state with spin projection $M=0$ ($M=1/2$) in an even-mass (odd-mass) nucleus. Denoting the $M$-projected density by $\rho_M$, the level density $\tilde{\rho}$ is given by
\be\label{level-density}
\tilde{\rho}  = \left \lbrace \begin{array}{lc} \rho_{M=0} & \mbox{for even-mass nuclei} \\
                                                   \rho_{M=1/2} & \mbox{for odd-mass nuclei} \end{array} \right . \; .
\ee
Projection on the spin component $M$ can be carried out by a discrete Fourier transform~\cite{Alhassid2007}. For example, the partition function of the propagator $U_\sigma$ at a fixed $M$ is given by
\begin{equation}
  \label{M-project}
  {\rm Tr}_M U_\sigma = {1 \over 2J_s + 1} \sum\limits_{k =-J_s}^{J_s}
  e^{-i \varphi_k M} {\rm Tr} \left( e^{i\varphi_k \hat J_z} U_\sigma \right) \;,
\end{equation}
where $\varphi_k$ ($k=-J_s,\ldots, J_s$) are quadrature points $\varphi_k=\pi {k \over J_s+1/2}$ and $J_s$ is the maximal spin in the many-particle shell model space. In SMMC, we calculate the thermal energy $E_M(\beta)$ at fixed $M$ and then integrate the thermodynamic relation $-d\ln Z_M /d\beta = E_M(\beta)$ to find the partition function $Z_M(\beta)$. The density $\rho_M$ is calculated in the saddle-point approximation in analogy with the state density (see Sect.~\ref{state_densities}). We have
\be\label{eq:rho_M}
\rho_M  \approx {1 \over \sqrt{2\pi T^2 C_M}} e^{S_M} \;,
\ee
where $S_M$ and $C_M$ are, respectively, the $M$-projected canonical entropy and heat capacity.

\section{Application to nickel isotopes}
\label{nickel}

Recent experiments have determined the level densities of a family of nickel isotopes $^{59-64}$Ni by measuring proton evaporation spectra~\cite{Voinov2012}. We have used the method of Sect.~\ref{level_densities} to calculate microscopically the level densities of these nickel isotopes~\cite{Bonett2013}. In these calculations we have used the complete $pfg_{9/2}$ model space with the Hamiltonian of Ref.~\cite{Nakada1997}.

\subsection{Ground-state energies}
In order to compare the calculated level densities with experiments, it is necessary to determine the excitation energy $E_x=E-E_0$, where $E_0$ is the ground-state energy. Thus it is important to determine  accurately the ground-state energy.

\begin{figure}[h!]
\centering
%\sidecaption
\includegraphics[width=0.9\columnwidth,clip]{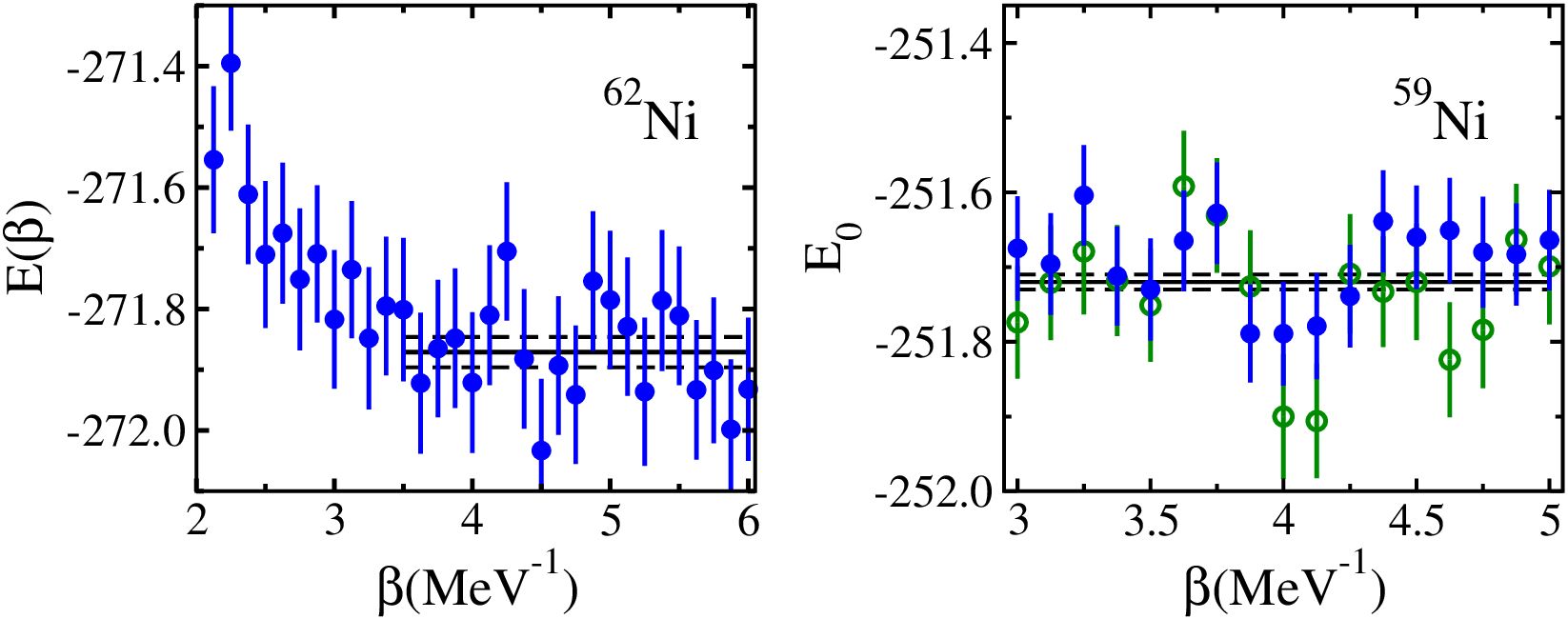}
\caption{Left: thermal energy $E(\beta)$ of the even-even nucleus $^{62}$Ni versus $\beta$. The solid horizontal line is our estimate for the ground-state energy $E_0$, which is determined by taking an average of $E(\beta)$ over the $\beta$ values within the length of the line, and the dashed lines indicate the error bar of $E_0$. Our estimate for the ground-state energy of $^{62}$Ni is $E_0= -271.87(3)$ MeV.
Right: the ground-state energy $E_0$ of $^{59}$Ni versus $\beta$, as determined from the Green's functions of $^{58}$Ni (open circles) and $^{60}$Ni (solid circles). The solid horizontal line describes the final value $E_0$, determined as an average over the length of the line and over both the $^{58}$Ni and $^{60}$Ni results. The dashed lines indicate the error bar of $E_0$. Our estimate for the ground-state energy of $^{59}$Ni is $E_0=  -251.72(1)$ MeV. }
\label{Ni62_59_gs}       
\end{figure}

For even-even nuclei, we have used two methods to determine $E_0$~\cite{Bonett2013}. In the first method, we simply calculate $E(\beta)$ for large values of $\beta$, where the contribution of the lowest excited $2^+$ state is negligible, and take an average. This is demonstrated in the left panel of Fig.~\ref{Ni62_59_gs}, where the SMMC thermal energy $E(\beta)$ of $^{62}$Ni is plotted versus $\beta$.  Above $\sim 3.5$ MeV$^{-1}$, $E(\beta)$ saturates and the ground-state energy $E_0$ is calculated as the average of $E(\beta)$ over the interval in $\beta$  between $3.5$ MeV$^{-1}$  and $6$ MeV$^{-1}$. The solid line in Fig.~\ref{Ni62_59_gs} is the estimated ground-state energy and the size of the statistical error is indicated by the dashed lines. The lines extend over the interval where the average is taken.  

A second method is based on a two-level model~\cite{Nakada1998}, in which we assume that only the $J=0^+$ ground state and the first $2^+$ excited state contribute to the expectation values of observables at sufficiently large values of $\beta$. In this model $E(\beta)$ depends on both $E_0$ and the excitation energy $E_x^{2^+}$ of the $2^+$ level. We determine $E_x^{2^+}$ from the SMMC calculations of $\langle \mathbf{J}^2 \rangle$, and then find $E_0$ from $E(\beta)$ and the already known value of $E_x^{2^+}$~\cite{Bonett2013}.

We determined the ground-state energies of the odd-mass nickel isotopes using the Green's function method of Sect.~\ref{Green_functions}. Results for $^{59}$Ni are shown in the right panel of Fig.~\ref{Ni62_59_gs} using the Green's functions of $^{58}$Ni (open circles) and $^{60}$Ni (solid circles). Energies were obtained using the slopes of the corresponding Green's functions and averaged over two values of the time slice $\Delta\beta = 1/32, 1/64$ MeV$^{-1}$.

\subsection{Level densities}

\begin{figure}[h!]
\centering
\includegraphics[width=0.9\columnwidth,clip]{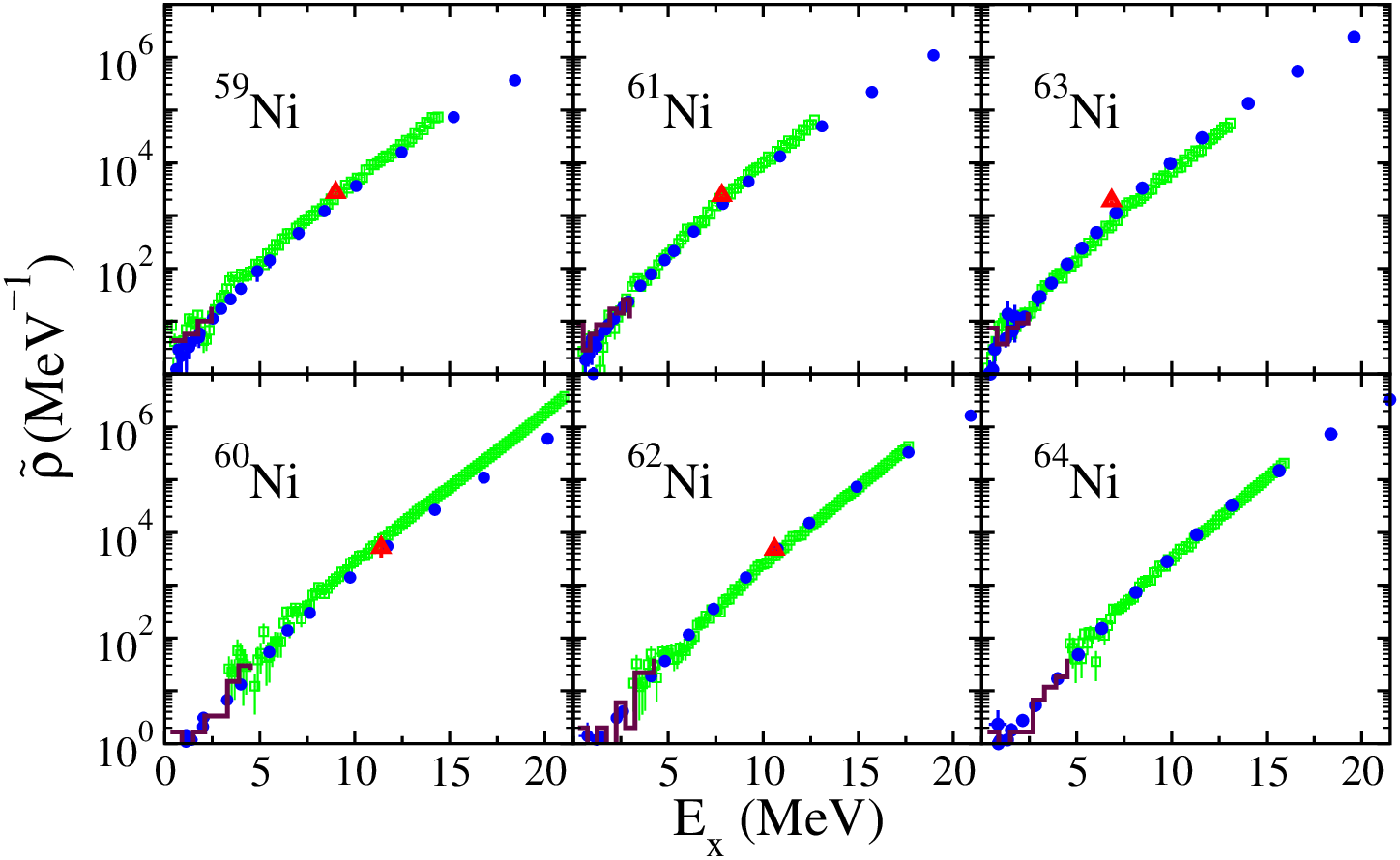}
\caption{Level densities for the nickel isotopes  $^{59-64}$Ni. The SMMC level densities (solid circles) are compared with level densities extracted from level counting at low excitation energies (solid histograms)~\cite{Ripl3}, proton evaporation spectra (open squares and quasi-continuous lines))~\cite{Voinov2012}, and neutron resonance data (triangles)~\cite{Ripl3} when available. We observe close agreement between theory and experiment. Taken from Ref.~\cite{Bonett2013}.
 }
\label{Ni_level}    
\end{figure}

We used Eqs.~(\ref{level-density}) and (\ref{eq:rho_M}) to calculate the level densities of the nickel isotopes. The results are shown in Fig.~\ref{Ni_level}, where the SMMC level densities (solid circles) are compared with level densities that are determined from several experimental data sets: (i)  direct level counting (solid histograms)~\cite{Ripl3}, (ii) proton evaporation spectra (open squares that merge into quasicontinuous lines)~\cite{Voinov2012} and (iii) neutron resonance data (triangle)~\cite{Ripl3}.

The level counting data take into account a complete set of experimental levels below a certain threshold. The proton evaporation spectra were measured recently by the Ohio University group in $^{6,7}$Li induced reactions on $^{54,56,58}$Fe. Level densities extracted from fits to the proton evaporation spectra were renormalized to the level counting data at low excitation energies. The level density at the neutron binding energy is obtained from the mean level spacing of $s$-wave resonances assuming parity equilibration and a spin cutoff model~\cite{Ericson1960} for the spin distribution with a rigid-body moment of inertia.  With the exception of the neutron resonance data in $^{63}$Ni, we observe close agreement between theory and experiment.

\subsection{Pairing correlations}

Pairing correlations lead in the thermodynamic limit to a phase transition to a superconducting phase below a certain critical temperature. In a finite-size system such as the nucleus there are no phase transitions, and an interesting question is whether signatures of the phase transition to a superconductor survive. In a superconductor, the magnetic susceptibility, which measures the response to an external magnetic field, is suppressed below the critical temperature. An analogous quantity in the nucleus is the moment of inertia,   which measures the response to rotations.

\begin{figure}[b]
\centering
\includegraphics[width=0.8\columnwidth,clip]{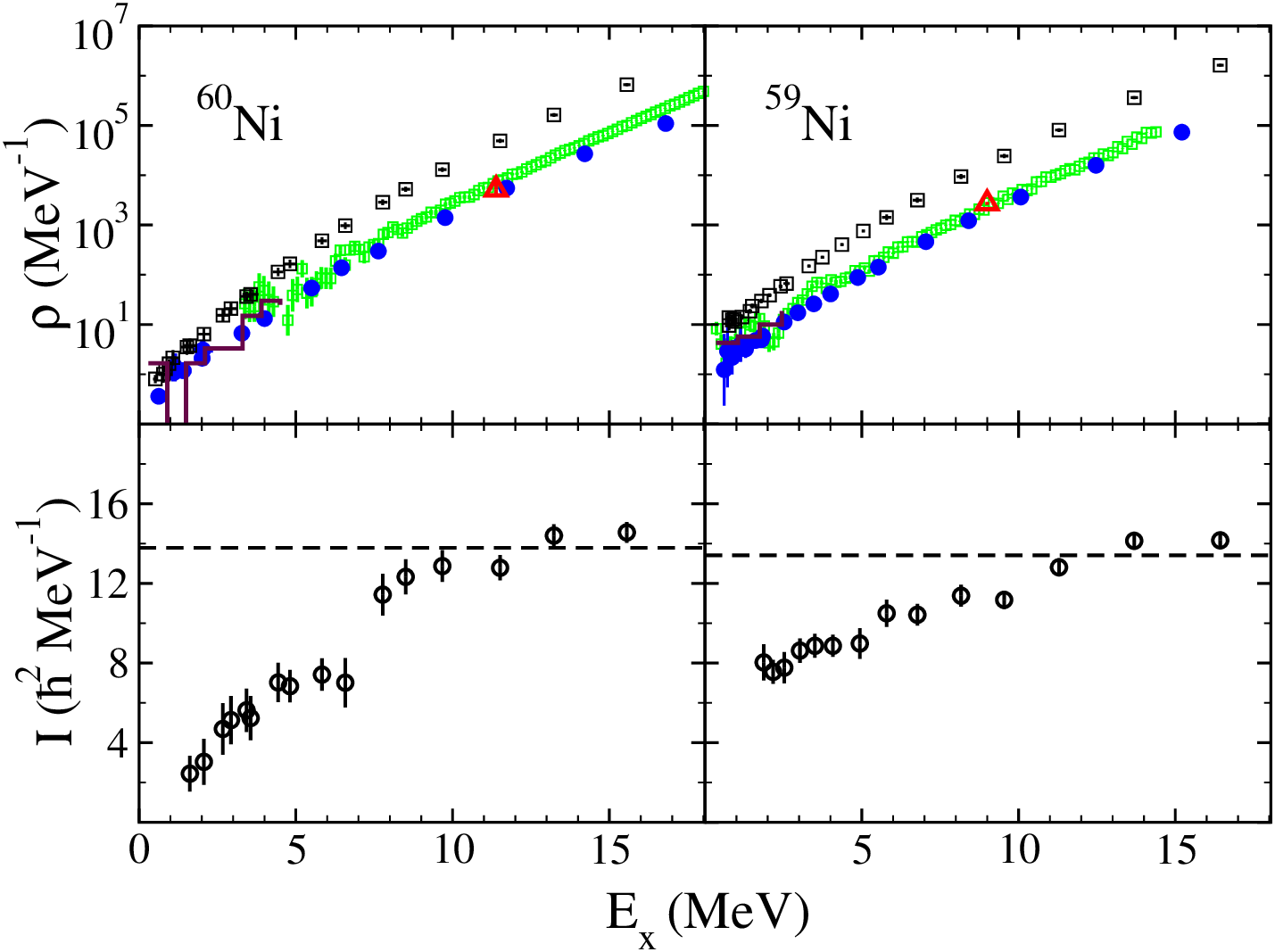}
\caption{Top panels: SMMC state density $\rho(E_x)$ (open squares) and level density $\tilde\rho(E_x)$ (solid circles) for $^{60}$Ni (left) and $^{59}$Ni (right). Experimental level densities are also shown, using the same notation as in Fig.~\ref{Ni_level}.
Bottom panels: moment of inertia $I$ versus excitation energy $E_x$ for $^{60}$Ni (left) and $^{59}$Ni (right). The dashed lines indicate the rigid-body values of the moment of inertia. The suppression of $I$ at low excitation energies is stronger in the even-even nucleus $^{60}$Ni than in the even-odd nucleus $^{59}$Ni.
 }
\label{Ni60_59_I}    
\end{figure}

We can extract a thermal moment of inertia from the calculated state and level densities as follows. In the spin cutoff model~\cite{Ericson1960}, the partial level density $\rho_J(E_x)$ at spin $J$ is given by
 \be
  \label{spin-cutoff}
  \rho_J(E_x) = \rho(E_x)
  {(2J+1) \over 2\sqrt{2 \pi} \sigma_c^3} e^{-{J(J+1) \over 2
      \sigma_c^2}}\;,
\ee
where $\sigma_c$ is a spin cutoff parameter that depends on the excitation energy $E_x$, and $\rho(E_x)$ is the total state density. The spin densities satisfy the normalization condition $\sum_J (2J+1) \rho_J(E_x) \approx \rho(E_x)$, while the level density can be calculated from
$\tilde \rho(E_x) = \sum_J \rho_J(E_x) \approx (2\pi)^{-1/2}\, \sigma_c^{-1} \rho(E_x)$ (where we have approximated the sum by an integral).  Thus, the spin cutoff parameter $\sigma_c$ can be extracted directly from the ratio between the state density and the level density
\be
\sigma_c(E_x) = (2\pi)^{-1/2}{ \rho(E_x) \over \tilde\rho(E_x)} \;.
\ee
The moment of inertia I is then determined from the relation
\be
\sigma_c^2 = {I T \over \hbar^2} \;.
\ee

In Fig.~\ref{Ni60_59_I} we show the SMMC state and level densities of the even-mass $^{60}$Ni (top left panel) and of the odd-mass nucleus $^{59}$Ni (top right panel).  The bottom panels show the extracted moment of inertia $I$ versus excitation energy for $^{60}$Ni (left) and $^{59}$Ni (right). In the even-even nucleus $^{60}$Ni we observe a strong suppression of the moment of inertia at low excitation energies relative with the rigid-body moment of inertia (dashed lines), while in the even-odd nucleus $^{59}$Ni  this suppression is weaker.  This odd-even effect, observed at excitation energies below $\sim 8$ MeV, is a clear signature of pairing correlations in the finite nucleus.

\section{Application to $^{162}$Dy}
In this section we apply the method of Sect.~\ref{level_densities} to calculate the level density of a typical heavy deformed rare-earth nucleus  $^{162}$Dy~\cite{Alhassid2013}. The single-particle model space consists of the $50-82$ major shell plus the $1f_{7/2}$ orbital for protons and the $80-126$ major shell plus the $0h_{11/2}$ and $1g_{9/2}$ orbitals for neutrons. We use the Hamiltonian of Ref.~\cite{Alhassid2008}.

\begin{figure}[h!]
\centering
\sidecaption
\includegraphics[width=0.6\columnwidth,clip]{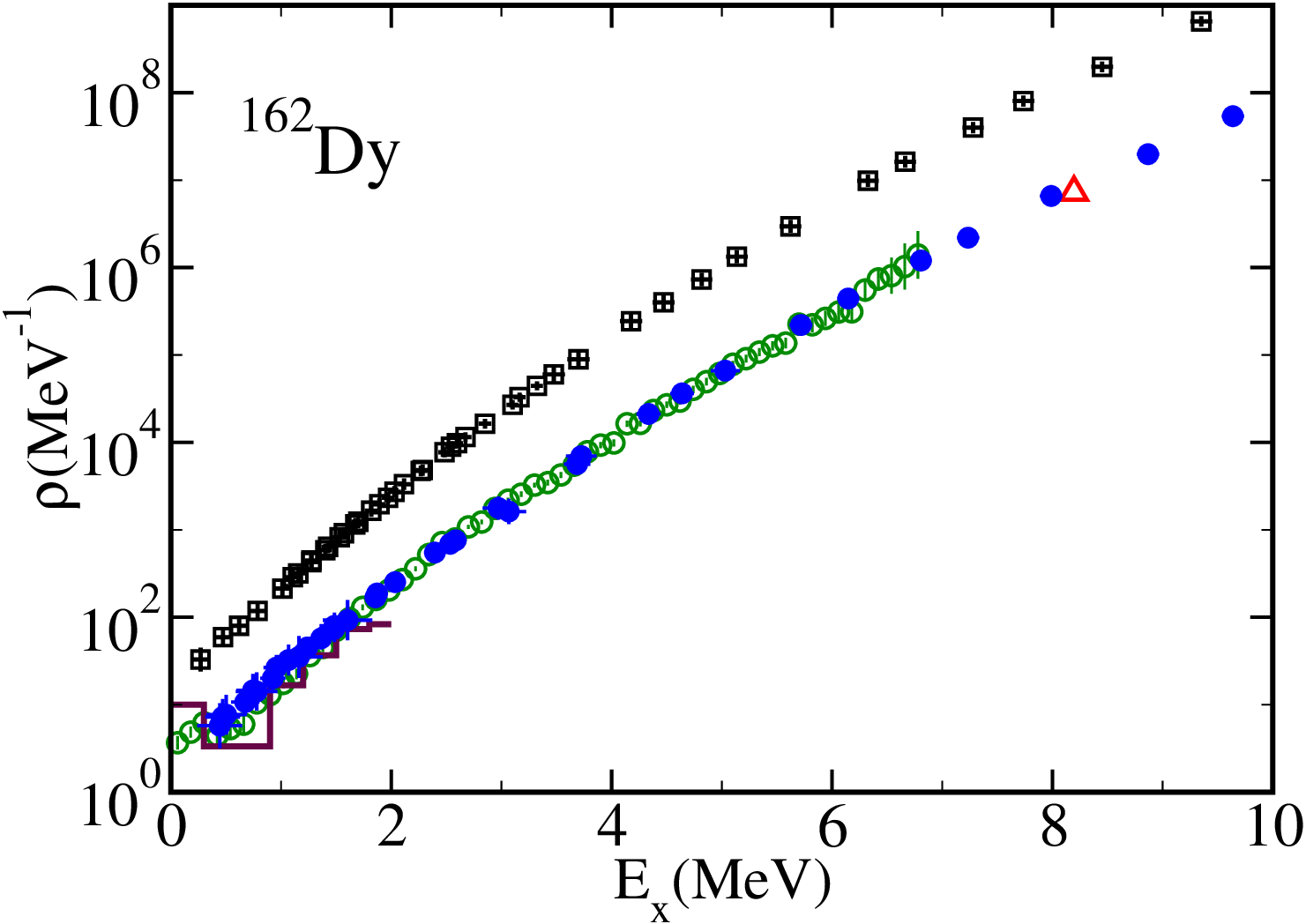}
\caption{SMMC level density (solid circles) and state density (open squares) for $^{162}$Dy. The SMMC level density is in good agreement with level counting data (solid histograms), renormalized Oslo data (open circles)~\cite{Oslo2000,Guttormsen2003}, and neutron resonance data (triangle)~\cite{Ripl3}. Taken from Ref.~\cite{Alhassid2013}.}
\label{Dy162_level}    
\end{figure}

In Fig.~\ref{Dy162_level} we show the SMMC level density (solid circles) of $^{162}$Dy in comparison with its SMMC state density (open squares). The SMMC level density is in good agreement with complete level counting data at excitation energies below $E_x\sim 2$ MeV (solid histograms), renormalized Oslo data (open circles)~\cite{Oslo2000,Guttormsen2003} and neutron resonance data (triangle)~\cite{Ripl3}.

\section{Conclusion}
The SMMC is a powerful method for the microscopic calculation of nuclear state densities in the presence of correlations. We have reviewed two recent technical developments and their applications.  One development is a method to calculate accurate ground-state energies of odd-mass nuclei despite a sign problem that originates from the projection on an odd number of particles. This method allows us to calculate accurate densities of odd-mass nuclei. The second development is a spin-projection method to calculate directly level densities, in which each level is counted once irrespective of its magnetic degeneracy.

Using these new methods, we calculated level densities in a family of nickel isotopes $^{59-64}$Ni (including the odd-mass isotopes) and the level density of a typical deformed heavy nucleus $^{162}$Dy. These calculated level densities can be compared directly with various sets of experimental data, and we find close agreement between theory and experiment.

This work was supported in part by the U.S. Department of Energy Grant No.~DE-FG02-91ER40608, and by the Grant-in-Aid for Scientific Research (C) No. 25400245 by the JSPS, Japan. Computational cycles were provided by the NERSC high performance computing facility at LBL and by the High Performance Computing Center at Yale University.

\end{document}